# Single-Molecule Magnets: Preparation and Properties of Mixed-Carboxylate Complexes [Mn$_{12}$O$_{12}$(O$_2$CR)$_8$(O$_2$CR′)$_8$(H$_2$O)$_4$]

Mònica Soler,[1a] Pau Artus,[1a] Kirsten Folting,[1a] John C. Huffman,[1a] David N. Hendrickson,*,[1b] and George Christou*,†,[1a]

*Department of Chemistry and Molecular Structure Center, Indiana University, Bloomington, Indiana 47405-7102, and Department of Chemistry-0358, University of California at San Diego, La Jolla, California 92093*

*Received April 16, 2001*

Methods are reported for the preparation of mixed-carboxylate versions of the [Mn$_{12}$O$_{12}$(O$_2$CR)$_{16}$(H$_2$O)$_4$] family of single-molecule magnets (SMMs). [Mn$_{12}$O$_{12}$(O$_2$CCHCl$_2$)$_8$(O$_2$CCH$_2$Bu$^t$)$_8$(H$_2$O)$_3$] (**5**) and [Mn$_{12}$O$_{12}$(O$_2$CHCl$_2$)$_8$-(O$_2$CEt)$_8$(H$_2$O)$_3$] (**6**) have been obtained from the 1:1 reaction of the corresponding homocarboxylate species. Complex **5**·CH$_2$Cl$_2$·H$_2$O crystallizes in the triclinic space group $P\bar{1}$ with, at −165 °C, $a = 15.762(1)$, $b = 16.246(1)$, $c = 23.822(1)$ Å, $\alpha = 103.92(1)$, $\beta = 104.50(1)$, $\gamma = 94.23(1)°$, $Z = 2$, and $V = 5674(2)$ Å$^3$. Complex **6**·CH$_2$Cl$_2$ crystallizes in the triclinic space group $P\bar{1}$ with, at −158 °C, $a = 13.4635(3)$, $b = 13.5162(3)$, $c = 23.2609(5)$ Å, $\alpha = 84.9796(6)$, $\beta = 89.0063(8)$, $\gamma = 86.2375(6)°$, $Z = 2$, and $V = 4207.3(3)$ Å$^3$. Complexes **5** and **6** both contain a [Mn$_{12}$O$_{12}$] core with the CHCl$_2$CO$_2^−$ ligands ordered in the axial positions and the RCO$_2^−$ ligands (R = CH$_2$Bu$^t$ (**5**) or Et (**6**)) in equatorial positions. There is, thus, a preference for the CHCl$_2$CO$_2^−$ to occupy the sites lying on the Mn$^{III}$ Jahn−Teller axes, and this is rationalized on the basis of the relative basicities of the carboxylate groups. Direct current magnetic susceptibility studies in a 10.0 kG field in the 2.00−300 K range indicate a large ground-state spin, and fitting of magnetization data collected in the 10.0−70.0 kG field and 1.80−4.00 K temperature range gave $S = 10$, $g = 1.89$, and $D = −0.65$ K for **5**, and $S = 10$, $g = 1.83$, and $D = −0.60$ K for **6**. These values are typical of [Mn$_{12}$O$_{12}$(O$_2$CR)$_{16}$(H$_2$O)$_4$] complexes. Alternating current susceptibility studies show the out-of-phase susceptibility ($\chi_M''$) signals characteristic of the slow relaxation in the millisecond time scale of single-molecule magnets. Arrhenius plots obtained from $\chi_M''$ versus $T$ data gave effective barriers to relaxation ($U_{eff}$) of 71 and 72 K for **5** and **6**, respectively. $^1$H NMR spectra in CD$_2$Cl$_2$ show that **5** and **6** are the main species present on dissolution, but there is evidence for some ligand distribution between axial and equatorial sites, by intra- and/or intermolecular exchange processes.

## Introduction

In 1993, it was discovered that single molecules of [Mn$_{12}$O$_{12}$(O$_2$CMe)$_{16}$(H$_2$O)$_4$] (**1**) are superparamagnets and can thus function as magnets at temperatures lower than their blocking temperature ($T_B$).[2] Such molecules were subsequently named single-molecule magnets (SMMs).[3] There has since been great interest in understanding this new magnetic phenomenon and in discovering other molecular complexes that exhibit similar properties. Several new examples have indeed been found in the last several years, including more manganese,[4−7] vanadium,[8] and iron[9] species. All these molecules possess the combination of sufficiently large ground-state spin ($S$) and negative (easy axis) magnetoanisotropy required to lead to the slow relaxation of magnetization required for single-molecule magnetism behavior. However, of the SMMs known to date, the [Mn$_{12}$O$_{12}$(O$_2$CR)$_{16}$(H$_2$O)$_4$] (R = various) family possesses the best structural and electronic properties for this phenomenon, inasmuch as it displays the SMM behavior at the highest temperatures.

Our understanding of the physics of nanoscale magnetic particles has also benefited from the availability of SMMs, because the latter provided the first truly monodisperse nanomagnet. Thus, quantum effects that had been predicted but never clearly seen were finally observed, such as quantum tunneling of magnetization (QTM).[10,11] This has led to a rapid expansion of the field and helped our understanding of the consequences of taking magnetic materials to the limit of miniaturization, where quantum effects become significant.

† Present address: Department of Chemistry, University of Florida, Gainesville, Florida 32611-7200.
(1) (a) Indiana University. (b) University of California at San Diego.
(2) (a) Sessoli, R.; Tsai, H.-L.; Schake, A. R.; Wang, S.; Vincent, J. B.; Folting, K.; Gatteschi, D.; Christou, G.; Hendrickson, D. N. *J. Am. Chem. Soc.* **1993**, *115*, 1804. (b) Sessoli, R.; Gatteschi, D.; Caneschi, A.; Novak, M. A. *Nature* **1993**, *365*, 141.
(3) Aubin, S. M. J.; Wemple, M. W.; Adams, D. M.; Tsai, H.-L.; Christou, G.; Hendrickson, D. N. *J. Am. Chem. Soc.* **1996**, *118*, 7746.
(4) (a) Tsai, H.-L.; Eppley, H. J.; de Vries, N.; Folting, K.; Christou, G.; Hendrickson, D. N. *Chem. Commun.* **1994**, 1745. (b) Eppley, H. J.; Tsai, H.-L.; de Vries, N.; Folting, K.; Christou, G.; Hendrickson, D. N. *J. Am. Chem. Soc.* **1995**, *117*, 301.
(5) (a) Aubin, S. M. J.; Dilley, N. R.; Wemple, M. W.; Maple, M. B.; Christou, G.; Hendrickson, D. N. *J. Am. Chem. Soc.* **1998**, *120*, 839. (b) Aubin, S. M. J.; Dilley, N. R.; Pardi, L.; Krzystek, J.; Wemple, M. W.; Brunel, L.-C.; Maple, M. B.; Christou, G.; Hendrickson, D. N. *J. Am. Chem. Soc.* **1998**, *120*, 4991.
(6) (a) Aubin, S. M. J.; Sun, Z.; Guzei, I. A.; Rheingold, A. L.; Christou, G.; Hendrickson, D. N. *Chem. Commun.* **1997**, 2239. (b) Sun, Z.; Ruiz, D.; Rumberger, E.; Incarvito, C. D.; Folting, K.; Rheingold, A. L.; Christou, G.; Hendrickson, D. N. *Inorg. Chem.* **1998**, *37*, 4758.
(7) (a) Brechin, E. K.; Yoo, J.; Nakano, M.; Huffman, J. C.; Hendrickson, D. N.; Christou, G. *Chem. Commun.* **1999**, 783. (b) Yoo, J.; Brechin, E. K.; Yamaguchi, A.; Nakano, M.; Huffman, J. C.; Maniero, A. L.; Brunel, L.-C.; Awaga, K.; Ishimoto, H.; Christou, G.; Hendrickson, D. N. *Inorg. Chem.* **2000**, *39*, 3615.
(8) Castro, S. L.; Sun, Z.; Grant, C. M.; Bollinger, J. C.; Hendrickson, D. N.; Christou, G. *J. Am. Chem. Soc.* **1998**, *120*, 2365.
(9) (a) Sangregorio, C.; Ohm, T.; Paulsen, C.; Sessoli, R.; Gatteschi, D. *Phys. Rev. Lett.* **1997**, *78*, 4645. (b) Oshio, H.; Hoshino, N.; Ito, T. *J. Am. Chem. Soc.* **2000**, *122*, 12602. (c) Barra, A.-L.; Caneschi, A.; Cornia, A.; Fabrizi de Biani, F.; Gatteschi, D.; Sangregorio, C.; Sessoli, R.; Sorace, L. *J. Am. Chem. Soc.* **1999**, *121*, 5302. (d) Barra, A.-L.; Debrunner, P.; Gatteschi, D.; Schulz, C. E.; Sessoli, R. *Europhys. Lett.* **1996**, *35*, 133.

Important to the potential future use of SMMs in, for example, high-density information storage or quantum computing is the ability to modify such molecules in a controllable fashion, both to assess the influence on the magnetic properties and to allow their assembly into ordered arrays. Within the $S = 10$ $Mn_{12}$ area, methods have been developed to exchange the $MeCO_2^-$ ligands of **1** for almost any other type of carboxylate group, improving solubility in organic solvents and greatly altering redox properties, among other benefits.[4b] This led, for example, to their subsequent conversion to the one-[4b,11,12] and two-electron[13] reduced forms $[Mn_{12}O_{12}(O_2CR)_{16}(H_2O)_4]^{-,2-}$, with $S = 19/2$ and 10, respectively.

More recently, we have also shown that it is possible to selectively replace four of the $RCO_2^-$ groups with $NO_3^-$ groups, giving $[Mn_{12}O_{12}(NO_3)_4(O_2CR)_{12}(H_2O)_4]$ products that are the first to contain ligands other than carboxylates (and $H_2O$).[14] This functionalization should prove very useful for a variety of further reactivity studies and applications, such as binding the complex to other species or surfaces without loss of the intrinsic $Mn_{12}$ structure or magnetic properties.

In the present work, we report a further type of site-specific modification of the $Mn_{12}$ family of complexes. Methods have been developed that allow mixed-carboxylate $[Mn_{12}O_{12}(O_2CR)_8(O_2CR')_8(H_2O)_3]$ complexes to be prepared with each carboxylate type localized at specific sites on the molecule. We report the preparation and characterization of two such molecules and describe their resulting spectroscopic and magnetic properties, the latter confirming that the products retain SMM properties.

**Experimental Section**

**Compound Preparation.** All chemicals and solvents were used as received. All preparations and manipulations were performed under aerobic conditions. $[Mn_{12}O_{12}(O_2CMe)_{16}(H_2O)_4]$ (**1**),[2a,15] $[Mn_{12}O_{12}(O_2CCH_2Bu^t)_{16}(H_2O)_4]$ (**2**),[16] and $[Mn_{12}O_{12}(O_2CEt)_{16}(H_2O)_3]$ (**3**)[2a] were prepared as previously described.

**$[Mn_{12}O_{12}(O_2CCHCl_2)_{16}(H_2O)_4]$ (4).** A solution of complex **1** (2.0 g, 0.97 mmol) in $CH_3CN$ (50 mL) was treated with a solution of $HO_2CCHCl_2$ (2.56 mL, 31.0 mmol) in $CH_2Cl_2$ (25 mL). The solution was stirred overnight, and the solvent was removed in vacuo. Toluene (20 mL) was added to the residue, and the solution was again evaporated to dryness. The addition and removal of toluene was repeated two more times. The remaining solid was redissolved in $CH_2Cl_2$ (60 mL) and treated again with a solution of $HO_2CCHCl_2$ (2.56 mL, 31.0 mmol) in $CH_2Cl_2$ (15 mL). After 6 h, three more cycles of addition and removal of toluene were performed. The residue was redissolved in $CH_2Cl_2$ (25 mL) and filtered, an equal volume of hexanes added, and the solution kept in a refrigerator for 2 days. The resulting black crystals were collected by filtration, washed with hexanes, and dried in vacuo. The yield was 1.25 g (60%). Anal. Calcd (Found) for **4** ($C_{32}H_{24}O_{48}Cl_{32}Mn_{12}$): C, 12.94% (13.01%); H, 0.81% (0.99%).

**$[Mn_{12}O_{12}(O_2CCHCl_2)_8(O_2CCH_2Bu^t)_8(H_2O)_3]$ (5).** A solution of complex **2** (0.24 g, 0.086 mmol) in $CH_2Cl_2$ (10 mL) was treated with a solution of complex **4** (0.26 g, 0.086 mmol) in $CH_2Cl_2$ (10 mL). After 3 h, hexanes (50 mL) were added, the solution was filtered, and the filtrate was concentrated by evaporation under an air flow until precipitation had just commenced. The solution was then sealed and kept in a refrigerator for 2 or 3 days. The resultant black crystals were twice recrystallized from a mixture of $CH_2Cl_2$ (20 mL) and hexanes (20 mL). Well-formed black crystals were obtained, and these were collected by filtration, washed copiously with hexanes, and dried in vacuo. The yield was 18%. Anal. Calcd (Found) for **5**·$1/2C_6H_{14}$-($C_{67}H_{109}O_{47}Cl_{16}Mn_{12}$): C, 27.82% (28.08%); H, 3.80% (3.76%). A crystallography sample was grown slowly from $CH_2Cl_2$/hexanes and maintained in mother liquor to avoid solvent loss.

**$[Mn_{12}O_{12}(O_2CCHCl_2)_8(O_2CEt)_8(H_2O)_3]$ (6).** A solution of complex **3** (0.20 g, 0.096 mmol) in $CH_2Cl_2$ (10 mL) was added to a solution of complex **4** (0.29 g, 0.096 mmol) in $CH_2Cl_2$ (10 mL). After 5 h, the solution was filtered, hexanes (50 mL) were added to the filtrate, and the solution was stored for 3 days in a refrigerator. A microcrystalline product was obtained, which was recrystallized from a mixture of $CH_2Cl_2$ (20 mL) and hexanes (20 mL). The resultant black crystals were collected by filtration, washed with hexanes, and dried in vacuo. The yield was 25%. Anal. Calcd (Found) for **6**·$0.3C_6H_{14}$: C, 19.63% (19.74%); H, 2.37% (2.19%). A crystallography sample was grown slowly from $CH_2Cl_2$/hexanes and maintained in mother liquor to avoid solvent loss.

**X-ray Crystallography.** Suitable single crystals were selected, attached to glass fibers using silicone grease, and transferred to the goniostat where they were cooled for characterization and data collection. For both complexes, examination of a limited portion of reciprocal space indicated no Laue symmetry or systematic absences, and the choice of centrosymmetric space group $P\bar{1}$ was confirmed by the successful structure solution and refinement. Data were collected on a Bruker platform goniometer equipped with a SMART 6000 CCD detector. Frames were measured for 5 s each with a frame width of 0.3° in $\omega$. Four sets of varying numbers of frames were measured, each set at a different $\varphi$ position. The frames were processed and integrated with the use of Bruker's SAINT software, and reflections with $6° < 2\theta < 55°$ were used. Data were additionally corrected for absorption effects (SADABS). The final averaging of redundant data was carried out using programs in the MSC XTEL software library.[17] The structures were solved by direct methods (XS in the Bruker software package).

For **5**·$CH_2Cl_2$·$H_2O$, all non-hydrogen atoms were readily located and refined. Hydrogen atoms were introduced as fixed contributors. Some disorder problems were encountered. Some $Bu^t$ groups showed rotational disorders of their Me groups, and some $CHCl_2$ groups showed rotational disorder of their Cl atoms. A $CH_2Cl_2$ solvent of crystallization was located, as well as an oxygen atom assigned to a $H_2O$ molecule. A final difference Fourier map was essentially featureless, the largest peaks being 1.45 e/Å$^3$ or less, generally in the vicinity of the disordered $O_2CCHCl_2$ groups.

For **6**·$CH_2Cl_2$, all non-hydrogen atoms were readily located. Hydrogen atoms were included as fixed contributors. Rotational disorder of $CHCl_2$ groups was again observed as well as a $O_2CCHCl_2$/$O_2CEt$ disorder at one equatorial carboxylate position, refining to 38%:62% partial occupancies, respectively. In addition, one axial $O_2CCHCl_2$ group is disordered with a neighboring $H_2O$ molecule (oxygen O(25)), with 75%:25% occupancies. The final difference Fourier map was relatively featureless: the largest peak was 2.4 e/Å$^3$ near C(56) of a $O_2CEt$ group, but attempts at refining a possible disorder at that site led to chemically unreasonable results. The next largest peaks were near the disordered $CHCl_2$ groups.

Crystallographic data for the two structures are listed in Table 1.

**Other Studies.** Infrared spectra were recorded on KBr pellets using a Nicolet 510P FTIR spectrophotometer. $^1H$ NMR spectra were obtained with a Varian XL-300 spectrometer; chemical shifts are quoted using the $\delta$ scale (shifts downfield are positive). Variable-temperature magnetic susceptibility data were obtained with a Quantum Design MPMS-XL SQUID susceptometer at Indiana University equipped with a 7 T magnet. The microcrystalline samples were restrained in eicosane to prevent torquing. Pascal's constants were used to estimate the diamagnetic correction, which was subtracted from the experimental susceptibility to give the molar magnetic susceptibility ($\chi_M$).

---

(10) Friedman, J. R.; Sarachik, M. P. *Phys. Rev. Lett.* **1996**, *76*, 3830.
(11) (a) Thomas, L.; Lionti, F.; Ballou, R.; Gatteschi, D.; Sessoli, R.; Barbara, B. *Nature* **1996**, *383*, 145. (b) Tejada, J.; Ziolo, R. F.; Zhang, X. X. *Chem. Mater.* **1996**, *8*, 1784.
(12) (a) Aubin, S. M. J.; Spagna, S.; Eppley, H. J.; Sager, R. E.; Christou, G.; Hendrickson, D. N. *Chem. Commun.* **1998**, 803. (b) Aubin, S. M. J.; Spagna, S.; Eppley, H. J.; Sager, R. E.; Christou, G.; Hendrickson, D. N. *Chem. Commun.* **1998**, 803.
(13) Soler, M.; Chandra, S. K.; Ruiz, D.; Davidson, E. R.; Hendrickson, D. N.; Christou, G. *Chem. Commun.* **2000**, 2417.
(14) Artus, P.; Boskovic, C.; Yoo, J.; Streib, W. E.; Brunel, L.-C.; Hendrickson, D. N.; Christou, G. Submitted for publication.
(15) Lis, T. *Acta Crystallogr.* **1980**, *B36*, 2042.
(16) Artus, P.; Christou, G. Manuscript in preparation.
(17) Huffman, J. C.; Bollinger, J. C.; Folting, K.; Streib, W. E. *XTEL, Computer Program Library*, Indiana University Molecular Structure Center, Bloomington, IN 47405, unpublished.

**Table 1.** Crystallographic Data for
[$Mn_{12}O_{12}(O_2CCHCl_2)_8(O_2CCH_2Bu^t)_8(H_2O)_3$]·$CH_2Cl_2$·$H_2O$
(**5**·$CH_2Cl_2$·$H_2O$) and [$Mn_{12}O_{12}(O_2CCHCl_2)_8(O_2CEt)_8(H_2O)_3$]·$CH_2Cl_2$
(**6**·$CH_2Cl_2$)

| | 5 | 6 |
|---|---|---|
| formula[a] | $C_{65}H_{106}Cl_{18}Mn_{12}O_{48}$ | $C_{41}H_{56}Cl_{18}Mn_{12}O_{47}$ |
| fw, g/mol | 2952.93 | 2598.27 |
| cryst syst | triclinic | triclinic |
| space group | $P\bar{1}$ | $P\bar{1}$ |
| $a$, Å | 15.762(1) | 13.4635(3) |
| $b$, Å | 16.246(1) | 13.5162(3) |
| $c$, Å | 23.822(1) | 23.2609(5) |
| $\alpha$, deg | 103.92(1) | 84.9796(6) |
| $\beta$, deg | 104.50(1) | 89.0063(8) |
| $\gamma$, deg | 94.23(1) | 86.2375(6) |
| $V$, Å$^3$ | 5674(2) | 4207.3(3) |
| $Z$ | 2 | 2 |
| cryst dimensions, mm | 0.45 × 0.40 × 0.40 | 0.30 × 0.40 × 0.40 |
| $T$, °C | −165 | −158 |
| radiation,[b] Å | 0.71073 | 0.71073 |
| $\rho_{calc}$, g/cm$^3$ | 1.729 | 2.051 |
| $\mu$, cm$^{-1}$ | 17.905 | 23.983 |
| total data | 40731 | 40702 |
| unique data | 25781 | 19274 |
| $R_{merge}$ | 0.045 | 0.029 |
| obsd data | 4549[c] | 13764[d] |
| $R(R_w)^{e,f}$ | 0.055(0.053) | 0.069(0.084) |

[a] Including solvent of crystallization. [b] Graphite monochromator. [c] $I > 2\sigma(I)$. [d] $I > 3\sigma(I)$. [e] $R = \Sigma||F_o| - |F_c||/\Sigma|F_o|$. [f] $R_w = [\Sigma w(|F_o| - |F_c|)^2/\Sigma w|F_o|^2]^{1/2}$ where $w = 1/\sigma^2(|F_o|)$.

**Scheme 1**

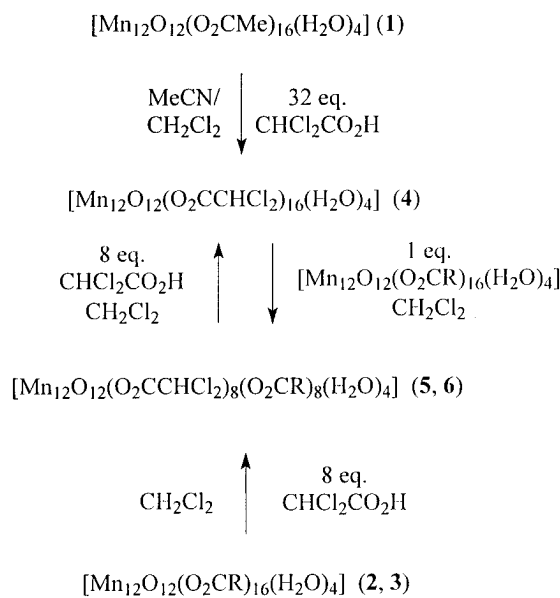

## Results and Discussion

**Ligand Substitution Reactions.** For convenience, the transformations described below are summarized in Scheme 1. One of the first reactivity characteristics of the [$Mn_{12}O_{12}(O_2CMe)_{16}$-$(H_2O)_4$] complex to be established was its ability to undergo carboxylate substitution reactions.[2a,4b] This provided an extremely useful and convenient means of accessing other carboxylate analogues, which provided benefits such as tunability of redox properties and increased solubility in a variety of organic solvents. However, the ligand substitution reaction is an equilibrium (eq 1) that must be driven to completion if pure product is to be obtained. Use of a RCO$_2$H that is more acidic than MeCO$_2$H ensures that the equilibrium favors the right-hand side, but even an excess of a more acidic RCO$_2$H does not guarantee *pure* product, as shown by the isolation of [$Mn_{12}O_{12}(O_2CMe)_3(O_2CPh)_{13}(H_2O)_4$] from the reaction of complex **1** with 32 equiv of PhCO$_2$H.[2a] Two methods have since been shown, however, to give pure product reliably: (i) isolation of the incompletely substituted product and then its redissolution and retreatment with another large excess of RCO$_2$H, and/or (ii) removal of MeCO$_2$H as its toluene azeotrope to drive the equilibrium of eq 1 to the right.[4b] The latter has proven particularly useful for the replacement of MeCO$_2^-$ with RCO$_2^-$ whose conjugate acid has a comparable or even greater p$K_a$ than MeCO$_2$H.[4b]

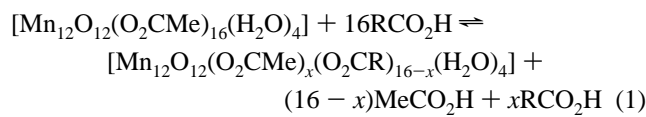

Various approaches were considered for the preparation of mixed-carboxylate complexes in the present work, and it was found that successful preparation of the desired materials could be accomplished most conveniently by the 1:1 reaction of [$Mn_{12}O_{12}(O_2CR)_{16}(H_2O)_4$] and [$Mn_{12}O_{12}(O_2CR')_{16}(H_2O)_4$] (eq 2). Ligand redistribution takes place, and the isolated product is the 8:8 complex [$Mn_{12}O_{12}(O_2CR)_8(O_2CR')_8(H_2O)_4$]. The RCO$_2^-$ and R'CO$_2^-$ groups are not disordered but instead are localized at specific sites on the molecule (vide infra). For unambiguous identification of R and R', and to ensure high solubility of the product, the Bu$^t$CH$_2$CO$_2^-$, EtCO$_2^-$, and CHCl$_2$CO$_2^-$ groups were chosen as representative carboxylates. The new CHCl$_2$CO$_2^-$ complex **4** was prepared from **1** by treatment with an excess of CHCl$_2$CO$_2$H, followed by removal of MeCO$_2$H as its azeotrope with toluene. The reaction of **4** with 1 equiv of [$Mn_{12}O_{12}(O_2CR)_{16}(H_2O)_4$] (R = CH$_2$Bu$^t$ (**2**) or Et(**3**)) led to the formation of mixed-carboxylate complexes **5** and **6** (eq 2) isolated in 18−25% yields after recrystallization. The high solubility imparted by Bu$^t$CH$_2$CO$_2^-$ and EtCO$_2^-$ groups is the cause of the low yields; we concentrated on obtaining the products as pure, well-formed crystals rather than optimizing overall yield.

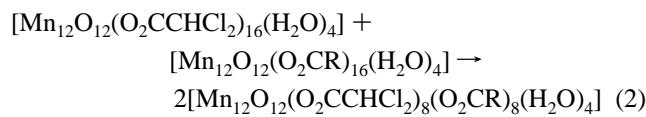

The crystal structures of **5** and **6** (vide infra) show that the CHCl$_2$CO$_2^-$ groups are localized in axial positions and the Bu$^t$CH$_2$CO$_2^-$ or EtCO$_2^-$ groups are in equatorial positions. This can be rationalized on the basis of their relative basicities. The acid p$K_a$ values are 1.48 (CHCl$_2$), 4.86 (Et), and 5.24 (CH$_2$-Bu$^t$); thus, the order of basicities of the RCO$_2^-$ groups is Bu$^t$-CH$_2$CO$_2$ > EtCO$_2^-$ ≫ CHCl$_2$CO$_2^-$. Sixteen of the twenty axial sites lie on Mn$^{III}$ Jahn−Teller (JT) elongation axes (the other axial sites are on Mn$^{IV}$ ions) and thus are lengthened by 0.1−0.2 Å and weakened relative to equatorial (non-JT-elongated) Mn−carboxylate bonds. This rationalizes why the less basic CHCl$_2$CO$_2^-$ groups are found at axial positions, because the more basic RCO$_2^-$ groups will favor equatorial sites where Mn−O bonds are shorter and bond energies can be maximized, increasing the stability of the whole molecule. Equatorial Mn−$O_2CCHCl_2$ bonds would be weaker, and axial Mn−$O_2CR$ bonds would still be JT elongated and weak. Thus, there is a small but, we believe, real preference for different carboxylates to occupy specific sites at the $Mn_{12}$ complex. This was an additional advantage of choosing complex **4** as one component of the reaction in eq 2 and is no doubt one of the main reasons that complexes **5** and **6** could be obtained in high isomeric purity vis-à-vis $CHCl_2CO_2^-/RCO_2^-$ disorder. We have not explored the reaction between complexes containing carboxylates with comparable basicities (e.g., $Bu^tCH_2$ vs Et, Me vs Et, etc.), anticipating that ligand disorder would be more of a problem. Note that some level of axial versus equatorial exchange is detected by NMR on redissolution of crystals of **5** or **6** (vide infra).

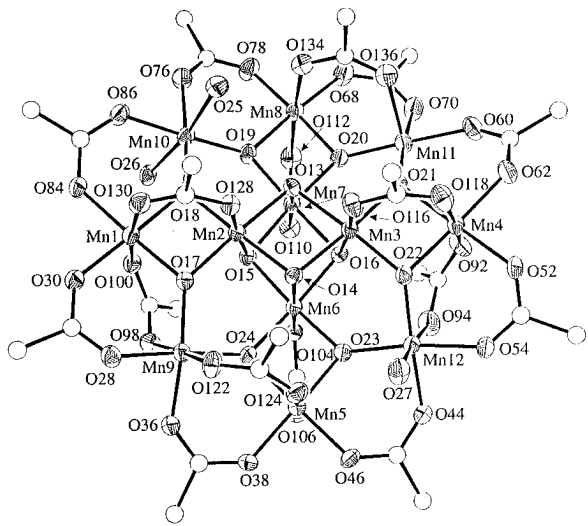

**Figure 1.** ORTEP representation of $[Mn_{12}O_{12}(O_2CCHCl_2)_8(O_2CCH_2Bu^t)_8(H_2O)_3]$ (**5**) at the 50% probability level. For clarity, only the α-C atom of the carboxylate R groups is shown.

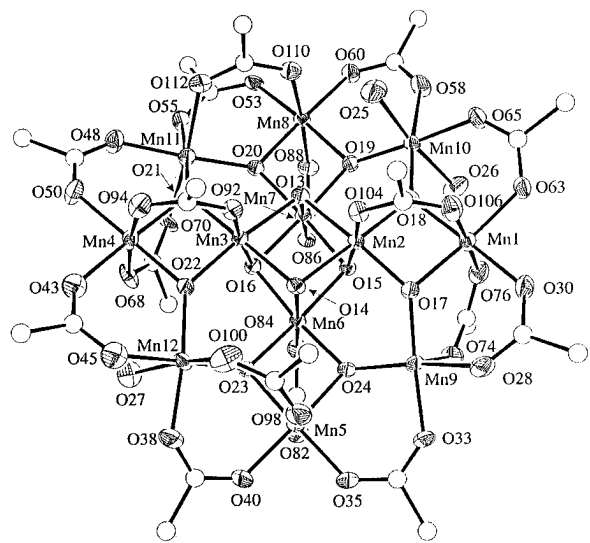

**Figure 2.** ORTEP representation of $[Mn_{12}O_{12}(O_2CCHCl_2)_8(O_2CEt)_8(H_2O)_4]$ (**6**) at the 50% probability level. For clarity, only the α-C atom of the carboxylate R groups is shown.

The large difference in p$K_a$ values between $CHCl_2CO_2H$ and $RCO_2H$ also provides an alternative route to **5** and **6**. Addition of 8 equiv of $CHCl_2CO_2H$ to **2** or **3** allows direct access to the mixed-carboxylate products (eq 3), as confirmed by spectroscopic examination of

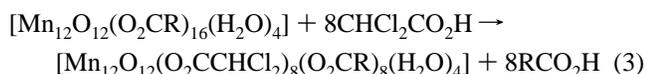

$$[Mn_{12}O_{12}(O_2CR)_{16}(H_2O)_4] + 8CHCl_2CO_2H \rightarrow$$
$$[Mn_{12}O_{12}(O_2CCHCl_2)_8(O_2CR)_8(H_2O)_4] + 8RCO_2H \quad (3)$$

the isolated solid. This is not a generally useful method, however, unless the leaving acid is preferentially removable by azeotropic distillation, because a smaller p$K_a$ difference between the incoming and leaving acids will yield an equilibrium mixture containing species with differing carboxylate compositions, as discussed above. Finally, treatment of **5** or **6** with 8 or more equiv of $CHCl_2CO_2H$ gives complex **4** (eq 4); the low p$K_a$ of $CHCl_2CO_2H$ causes essentially complete conversion to **4** even with only 8 equiv, as confirmed by $^1H$ NMR spectroscopy.

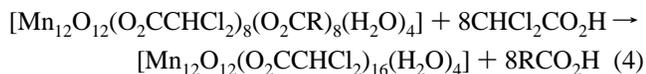

$$[Mn_{12}O_{12}(O_2CCHCl_2)_8(O_2CR)_8(H_2O)_4] + 8CHCl_2CO_2H \rightarrow$$
$$[Mn_{12}O_{12}(O_2CCHCl_2)_{16}(H_2O)_4] + 8RCO_2H \quad (4)$$

**Description of Structures.** Labeled ORTEP plots of **5** and **6** are shown in Figures 1 and 2, respectively. Selected bond distances and angles are listed in Tables 2 and 3.

Complex **5** crystallizes in the triclinic space group $P\bar{1}$ with the $Mn_{12}$ molecule in a general position. The complex has the same overall structure as other $[Mn_{12}O_{12}(O_2CR)_{16}(H_2O)_4]$ complexes,[2a,4b] consisting of a central $[Mn^{IV}_4O_4]$ cubane and an outer ring of eight $Mn^{III}$ ions. The latter are Jahn−Teller distorted with their JT elongation axes all axial with respect to the $Mn_{12}$ core. As occasionally observed previously,[4b] there are only three (rather than the usual four) $H_2O$ ligands (two on Mn(10) and one on Mn(12)), and one $Mn^{III}$ ion (Mn(11)) is consequently five-coordinate with square-pyramidal geometry ($\tau = 0.06$, where $\tau = 0$ and 1 for pure sp and tbp geometries, respectively[18]). This is almost certainly a solid-state effect, and in solution, it is anticipated that all $Mn^{III}$ ions are six-coordinate. (For this reason, four bound $H_2O$ molecules are assumed in discussing the preparative reactions and NMR spectra.)

Complex **6** also crystallizes in the triclinic space group $P\bar{1}$ with the $Mn_{12}$ molecule in a general position. The structure is overall very similar to that of complex **5**, and again, there are only three $H_2O$ molecules, with Mn(9) being five-coordinate and square-pyramidal. All $Mn^{III}$ JT axes are again axial to the $[Mn_{12}O_{12}]$ core.

For both complexes **5** and **6**, the $CHCl_2CO_2^-$ groups are axial and the $RCO_2^-$ groups are equatorial. Thus, the $CHCl_2CO_2^-$ groups are either bridging $Mn^{III}/Mn^{III}$ pairs with both their O atoms lying on the JT axes, or they are bridging $Mn^{III}/Mn^{IV}$ pairs with only one O atom on a JT elongation site. For complex **5** (Figure 1), these are the (Mn(1)/Mn(9), Mn(9)/Mn(5), Mn(12)/Mn(4), Mn(11)/Mn(8)) and (Mn(1)/Mn(2), Mn(5)/Mn(6), Mn(3)/Mn(4), Mn(7)/Mn(8)) pairs, respectively. The situation is similar for complex **6** (Figure 2). The eight equatorial $RCO_2^-$ groups of **5** and **6** bridge $Mn^{III}/Mn^{III}$ pairs with no atoms on JT elongation axes. Because the $Mn^{III}$ JT elongated (axial) sites are occupied by one type of carboxylate and the equatorial sites by another, it was of interest to determine if this had any structural consequence, particularly in the bond length difference between axial and equatorial $Mn^{III}$−O bonds. Listed in Table 4 are the averages for **5** and **6** of the axial and equatorial $Mn^{III}$−O bonds with respect to the JT elongation axes, together with a comparison with the same averages for homocarboxylate complexes **1**−**3**. The difference (Δ) between the axial and equatorial bond lengths is greater for **5** and **6** than for **1**−**3**.

---

(18) Jansen, J. C.; Van Koningsveld, H.; Van Ooijen, J. A. C.; Reedijk, J. *Inorg. Chem.* **1980**, *19*, 170.

**Table 2.** Selected Bond Distances (Å) and Angles (deg) for Complex **5**·CH$_2$Cl$_2$·H$_2$O

| | | | | | | | | | | | |
|---|---|---|---|---|---|---|---|---|---|---|---|
| Mn(1) | O(17) | 1.904(3) | Mn(4) | O(21) | 1.923(3) | Mn(7) | O(13) | 1.886(3) | Mn(10) | O(18) | 1.887(3) |
| Mn(1) | O(18) | 1.886(3) | Mn(4) | O(22) | 1.894(3) | Mn(7) | O(15) | 1.913(3) | Mn(10) | O(19) | 1.909(3) |
| Mn(1) | O(30) | 1.910(3) | Mn(4) | O(52) | 1.938(3) | Mn(7) | O(16) | 1.930(3) | Mn(10) | O(25) | 2.227(4) |
| Mn(1) | O(84) | 1.939(3) | Mn(4) | O(62) | 1.936(3) | Mn(7) | O(19) | 1.877(3) | Mn(10) | O(26) | 2.182(3) |
| Mn(1) | O(100) | 2.228(3) | Mn(4) | O(92) | 2.145(4) | Mn(7) | O(20) | 1.850(3) | Mn(10) | O(76) | 1.943(3) |
| Mn(1) | O(130) | 2.270(3) | Mn(4) | O(118) | 2.205(4) | Mn(7) | O(110) | 1.934(3) | Mn(10) | O(86) | 1.966(3) |
| Mn(2) | O(13) | 1.921(3) | Mn(5) | O(23) | 1.892(3) | Mn(8) | O(19) | 1.882(3) | Mn(11) | O(20) | 1.875(3) |
| Mn(2) | O(14) | 1.910(3) | Mn(5) | O(24) | 1.908(3) | Mn(8) | O(20) | 1.897(3) | Mn(11) | O(21) | 1.895(3) |
| Mn(2) | O(15) | 1.896(3) | Mn(5) | O(38) | 1.934(3) | Mn(8) | O(68) | 1.943(3) | Mn(11) | O(60) | 1.914(3) |
| Mn(2) | O(17) | 1.854(3) | Mn(5) | O(46) | 1.949(3) | Mn(8) | O(78) | 1.924(3) | Mn(11) | O(70) | 1.944(3) |
| Mn(2) | O(18) | 1.864(3) | Mn(5) | O(106) | 2.225(3) | Mn(8) | O(112) | 2.235(4) | Mn(11) | O(136) | 2.103(4) |
| Mn(2) | O(128) | 1.921(3) | Mn(5) | O(124) | 2.187(3) | Mn(8) | O(134) | 2.197(4) | Mn(12) | O(22) | 1.915(3) |
| Mn(3) | O(13) | 1.933(3) | Mn(6) | O(14) | 1.895(3) | Mn(9) | O(17) | 1.890(3) | Mn(12) | O(23) | 1.886(3) |
| Mn(3) | O(14) | 1.924(3) | Mn(6) | O(15) | 1.916(3) | Mn(9) | O(24) | 1.905(3) | Mn(12) | O(27) | 2.210(3) |
| Mn(3) | O(16) | 1.898(3) | Mn(6) | O(16) | 1.934(3) | Mn(9) | O(28) | 1.957(3) | Mn(12) | O(44) | 1.930(3) |
| Mn(3) | O(21) | 1.870(3) | Mn(6) | O(23) | 1.851(3) | Mn(9) | O(36) | 1.936(3) | Mn(12) | O(54) | 1.955(3) |
| Mn(3) | O(22) | 1.867(3) | Mn(6) | O(24) | 1.847(3) | Mn(9) | O(98) | 2.179(3) | Mn(12) | O(94) | 2.159(4) |
| Mn(3) | O(116) | 1.918(3) | Mn(6) | O(104) | 1.939(3) | Mn(9) | O(122) | 2.141(3) | | | |
| O(17) | Mn(1) | O(18) | 83.08(13) | O(38) | Mn(5) | O(124) | 93.77(14) | O(19) | Mn(10) | O(26) | 91.05(13) |
| O(17) | Mn(1) | O(30) | 96.19(13) | O(46) | Mn(5) | O(106) | 88.50(13) | O(19) | Mn(10) | O(76) | 91.88(14) |
| O(17) | Mn(1) | O(84) | 176.76(14) | O(46) | Mn(5) | O(124) | 94.94(13) | O(19) | Mn(10) | O(86) | 176.29(13) |
| O(17) | Mn(1) | O(100) | 91.86(13) | O(106) | Mn(5) | O(124) | 174.75(12) | O(25) | Mn(10) | O(26) | 171.58(12) |
| O(17) | Mn(1) | O(130) | 83.97(13) | O(14) | Mn(6) | O(15) | 83.49(13) | O(25) | Mn(10) | O(76) | 87.22(14) |
| O(18) | Mn(1) | O(30) | 175.49(14) | O(14) | Mn(6) | O(16) | 83.48(13) | O(25) | Mn(10) | O(86) | 84.83(14) |
| O(18) | Mn(1) | O(84) | 94.34(13) | O(14) | Mn(6) | O(23) | 90.49(13) | O(26) | Mn(10) | O(76) | 87.78(13) |
| O(18) | Mn(1) | O(100) | 90.58(12) | O(14) | Mn(6) | O(24) | 90.60(13) | O(26) | Mn(10) | O(86) | 87.96(13) |
| O(18) | Mn(1) | O(130) | 83.26(12) | O(14) | Mn(6) | O(104) | 174.93(13) | O(76) | Mn(10) | O(86) | 84.51(14) |
| O(30) | Mn(1) | O(84) | 86.22(14) | O(15) | Mn(6) | O(16) | 79.92(12) | O(20) | Mn(11) | O(21) | 92.91(12) |
| O(30) | Mn(1) | O(100) | 93.89(13) | O(15) | Mn(6) | O(23) | 173.36(14) | O(20) | Mn(11) | O(60) | 169.54(15) |
| O(30) | Mn(1) | O(130) | 92.23(13) | O(15) | Mn(6) | O(24) | 97.54(13) | O(20) | Mn(11) | O(70) | 90.21(14) |
| O(84) | Mn(1) | O(100) | 90.12(13) | O(15) | Mn(6) | O(104) | 92.58(13) | O(20) | Mn(11) | O(136) | 96.28(14) |
| O(84) | Mn(1) | O(130) | 93.79(13) | O(16) | Mn(6) | O(23) | 96.67(13) | O(21) | Mn(11) | O(60) | 95.18(14) |
| O(100) | Mn(1) | O(130) | 172.94(12) | O(16) | Mn(6) | O(24) | 173.77(14) | O(21) | Mn(11) | O(70) | 172.86(14) |
| O(13) | Mn(2) | O(14) | 79.95(12) | O(16) | Mn(6) | O(104) | 92.69(13) | O(21) | Mn(11) | O(136) | 95.82(14) |
| O(13) | Mn(2) | O(15) | 83.78(13) | O(23) | Mn(6) | O(24) | 85.28(13) | O(60) | Mn(11) | O(70) | 81.03(14) |
| O(13) | Mn(2) | O(17) | 174.07(14) | O(23) | Mn(6) | O(104) | 93.27(13) | O(60) | Mn(11) | O(136) | 89.48(15) |
| O(13) | Mn(2) | O(18) | 96.69(13) | O(24) | Mn(6) | O(104) | 93.11(13) | O(70) | Mn(11) | O(136) | 90.21(15) |
| O(13) | Mn(2) | O(128) | 92.53(13) | O(13) | Mn(7) | O(15) | 84.28(12) | O(22) | Mn(12) | O(23) | 93.43(13) |
| O(14) | Mn(2) | O(15) | 83.62(13) | O(13) | Mn(7) | O(16) | 84.18(13) | O(22) | Mn(12) | O(27) | 90.25(14) |
| O(14) | Mn(2) | O(17) | 97.67(13) | O(13) | Mn(7) | O(19) | 89.22(13) | O(22) | Mn(12) | O(44) | 177.23(13) |
| O(14) | Mn(2) | O(18) | 172.84(13) | O(13) | Mn(7) | O(20) | 92.60(13) | O(22) | Mn(12) | O(54) | 94.01(14) |
| O(14) | Mn(2) | O(128) | 94.04(13) | O(13) | Mn(7) | O(110) | 174.11(14) | O(22) | Mn(12) | O(94) | 91.49(14) |
| O(15) | Mn(2) | O(17) | 90.56(13) | O(15) | Mn(7) | O(16) | 80.08(12) | O(23) | Mn(12) | O(27) | 92.73(12) |
| O(15) | Mn(2) | O(18) | 89.75(13) | O(15) | Mn(7) | O(19) | 100.79(13) | O(23) | Mn(12) | O(44) | 89.33(14) |
| O(15) | Mn(2) | O(128) | 175.92(13) | O(15) | Mn(7) | O(20) | 174.01(13) | O(23) | Mn(12) | O(54) | 172.31(14) |
| O(17) | Mn(2) | O(18) | 85.05(13) | O(15) | Mn(7) | O(110) | 90.14(13) | O(23) | Mn(12) | O(94) | 91.04(13) |
| O(17) | Mn(2) | O(128) | 93.07(13) | O(16) | Mn(7) | O(19) | 173.23(13) | O(27) | Mn(12) | O(44) | 89.92(15) |
| O(18) | Mn(2) | O(128) | 92.42(14) | O(16) | Mn(7) | O(20) | 94.54(13) | O(27) | Mn(12) | O(54) | 85.28(13) |
| O(13) | Mn(3) | O(14) | 79.30(12) | O(16) | Mn(7) | O(110) | 93.03(14) | O(27) | Mn(12) | O(94) | 175.74(13) |
| O(13) | Mn(3) | O(16) | 83.78(13) | O(19) | Mn(7) | O(20) | 84.24(13) | O(44) | Mn(12) | O(54) | 83.25(14) |
| O(13) | Mn(3) | O(21) | 96.24(12) | O(19) | Mn(7) | O(110) | 93.68(14) | O(44) | Mn(12) | O(94) | 88.15(16) |
| O(13) | Mn(3) | O(22) | 174.70(14) | O(20) | Mn(7) | O(110) | 92.79(14) | O(54) | Mn(12) | O(94) | 90.73(14) |
| O(13) | Mn(3) | O(116) | 92.36(14) | O(19) | Mn(8) | O(20) | 82.81(13) | Mn(2) | O(13) | Mn(3) | 100.01(13) |
| O(14) | Mn(3) | O(16) | 83.65(12) | O(19) | Mn(8) | O(68) | 174.37(16) | Mn(2) | O(13) | Mn(7) | 95.10(13) |
| O(14) | Mn(3) | O(21) | 174.06(13) | O(19) | Mn(8) | O(78) | 95.74(14) | Mn(3) | O(13) | Mn(7) | 95.32(14) |
| O(14) | Mn(3) | O(22) | 98.30(12) | O(19) | Mn(8) | O(112) | 86.16(13) | Mn(2) | O(14) | Mn(3) | 100.70(13) |
| O(14) | Mn(3) | O(116) | 90.79(13) | O(19) | Mn(8) | O(134) | 94.98(13) | Mn(2) | O(14) | Mn(6) | 95.60(13) |
| O(16) | Mn(3) | O(21) | 92.01(13) | O(20) | Mn(8) | O(68) | 94.72(13) | Mn(3) | O(14) | Mn(6) | 95.71(13) |
| O(16) | Mn(3) | O(22) | 91.27(13) | O(20) | Mn(8) | O(78) | 175.14(15) | Mn(2) | O(15) | Mn(6) | 95.34(13) |
| O(16) | Mn(3) | O(116) | 173.73(13) | O(20) | Mn(8) | O(112) | 85.39(13) | Mn(2) | O(15) | Mn(7) | 95.01(13) |
| O(21) | Mn(3) | O(22) | 85.81(13) | O(20) | Mn(8) | O(134) | 92.17(13) | Mn(6) | O(15) | Mn(7) | 100.61(13) |
| O(21) | Mn(3) | O(116) | 93.33(14) | O(68) | Mn(8) | O(78) | 86.29(14) | Mn(3) | O(16) | Mn(6) | 95.29(13) |
| O(22) | Mn(3) | O(116) | 92.40(14) | O(68) | Mn(8) | O(112) | 88.60(15) | Mn(3) | O(16) | Mn(7) | 95.01(14) |
| O(21) | Mn(4) | O(22) | 83.60(12) | O(68) | Mn(8) | O(134) | 90.16(15) | Mn(6) | O(16) | Mn(7) | 99.38(13) |
| O(21) | Mn(4) | O(52) | 179.66(14) | O(78) | Mn(8) | O(112) | 89.89(14) | Mn(1) | O(17) | Mn(2) | 95.44(14) |
| O(21) | Mn(4) | O(62) | 95.39(13) | O(78) | Mn(8) | O(134) | 92.58(15) | Mn(1) | O(17) | Mn(9) | 124.65(16) |
| O(21) | Mn(4) | O(92) | 86.17(14) | O(112) | Mn(8) | O(134) | 177.16(13) | Mn(2) | O(17) | Mn(9) | 134.26(17) |
| O(21) | Mn(4) | O(118) | 86.75(13) | O(17) | Mn(9) | O(24) | 92.43(13) | Mn(1) | O(18) | Mn(2) | 95.69(13) |

**Table 2** (Continued)

| | | | | | | | | | | | |
|---|---|---|---|---|---|---|---|---|---|---|---|
| O(22) | Mn(4) | O(52) | 96.71(13) | O(17) | Mn(9) | O(28) | 92.21(14) | Mn(1) | O(18) | Mn(10) | 129.18(16) |
| O(22) | Mn(4) | O(62) | 174.27(15) | O(17) | Mn(9) | O(36) | 174.06(13) | Mn(2) | O(18) | Mn(10) | 133.06(16) |
| O(22) | Mn(4) | O(92) | 92.47(13) | O(17) | Mn(9) | O(98) | 89.97(13) | Mn(7) | O(19) | Mn(8) | 95.47(13) |
| O(22) | Mn(4) | O(118) | 85.78(13) | O(17) | Mn(9) | O(122) | 91.04(13) | Mn(7) | O(19) | Mn(10) | 131.55(16) |
| O(52) | Mn(4) | O(62) | 84.29(14) | O(24) | Mn(9) | O(28) | 175.36(14) | Mn(8) | O(19) | Mn(10) | 127.00(17) |
| O(52) | Mn(4) | O(92) | 93.95(15) | O(24) | Mn(9) | O(36) | 92.44(14) | Mn(7) | O(20) | Mn(8) | 95.83(14) |
| O(52) | Mn(4) | O(118) | 93.14(14) | O(24) | Mn(9) | O(98) | 90.20(13) | Mn(7) | O(20) | Mn(11) | 135.21(16) |
| O(62) | Mn(4) | O(92) | 93.09(14) | O(24) | Mn(9) | O(122) | 91.82(13) | Mn(8) | O(20) | Mn(11) | 123.89(16) |
| O(62) | Mn(4) | O(118) | 88.54(14) | O(28) | Mn(9) | O(36) | 82.93(14) | Mn(3) | O(21) | Mn(4) | 93.96(13) |
| O(92) | Mn(4) | O(118) | 172.85(14) | O(28) | Mn(9) | O(98) | 89.83(13) | Mn(3) | O(21) | Mn(11) | 133.59(16) |
| O(23) | Mn(5) | O(24) | 82.46(12) | O(28) | Mn(9) | O(122) | 88.07(13) | Mn(4) | O(21) | Mn(11) | 121.76(16) |
| O(23) | Mn(5) | O(38) | 174.81(14) | O(36) | Mn(9) | O(98) | 86.62(13) | Mn(3) | O(22) | Mn(4) | 95.02(14) |
| O(23) | Mn(5) | O(46) | 94.07(13) | O(36) | Mn(9) | O(122) | 92.19(14) | Mn(3) | O(22) | Mn(12) | 132.00(16) |
| O(23) | Mn(5) | O(106) | 84.58(13) | O(98) | Mn(9) | O(122) | 177.71(12) | Mn(4) | O(22) | Mn(12) | 123.07(16) |
| O(23) | Mn(5) | O(124) | 91.21(13) | O(18) | Mn(10) | O(19) | 93.49(13) | Mn(5) | O(23) | Mn(6) | 95.72(13) |
| O(24) | Mn(5) | O(38) | 95.99(13) | O(18) | Mn(10) | O(25) | 89.47(13) | Mn(5) | O(23) | Mn(12) | 129.32(16) |
| O(24) | Mn(5) | O(46) | 172.97(14) | O(18) | Mn(10) | O(26) | 94.89(13) | Mn(6) | O(23) | Mn(12) | 133.04(17) |
| O(24) | Mn(5) | O(106) | 85.09(13) | O(18) | Mn(10) | O(76) | 173.96(14) | Mn(5) | O(24) | Mn(6) | 95.30(13) |
| O(24) | Mn(5) | O(124) | 91.26(13) | O(18) | Mn(10) | O(86) | 90.16(13) | Mn(5) | O(24) | Mn(9) | 124.24(15) |
| O(38) | Mn(5) | O(46) | 86.93(14) | O(19) | Mn(10) | O(25) | 95.88(13) | Mn(6) | O(24) | Mn(9) | 134.50(17) |
| O(38) | Mn(5) | O(106) | 90.36(13) | | | | | | | | |

The variation appears significant and is consistent with the carboxylate basicity differences in **5** and **6** and the resulting relative bond strengths as discussed above; that is, the Mn–O(eq) and Mn–O(ax) bonds are on average slightly shorter and longer, respectively, for mixed carboxylates **5** and **6** compared with **1**–**3**.

The EtCO$_2^-$ group for **6** defined by O atoms O(43) and O(45) in Figure 2 showed evidence of disorder, and it was identified as an EtCO$_2^-$/CHCl$_2$CO$_2^-$ mixture which refined to 62%:38% relative occupancies, respectively. No corresponding mixed-carboxylate disorder was observed at an axial position, suggesting that the mixed occupancy at an equatorial position was not caused by EtCO$_2^-$/CHCl$_2$CO$_2^-$ exchange between axial and equatorial sites. Instead, we believe the 62%:38% disorder is because of the homocarboxylate Mn$_{12}$ starting materials not being in an exact 1:1 molar ratio for the reaction from which the crystallographic sample was taken, perhaps because of a change in solvation content with time that we did not realize when weighing out the reactants. A slight excess of **4** over **3** would necessarily give some product molecules with a CHCl$_2$CO$_2^-$–EtCO$_2^-$ ratio slightly greater than 1. No such disorder was observed for **5**.

**Magnetochemistry.** Variable-temperature dc magnetic susceptibility ($\chi_M$) data were collected on complexes **5** and **6** in the 2.00–300 K range in a 10 kG (1 T) magnetic field. The $\chi_M T$ versus $T$ dependences are similar to those of previously studied [Mn$_{12}$O$_2$(O$_2$CR)$_{16}$(H$_2$O)$_4$] complexes with $S = 10$ ground states, exhibiting a nearly temperature-independent value of 16–18 cm$^3$ K mol$^{-1}$ in the 150–300 K range which then increases rapidly to a maximum of 46–48 cm$^3$ K mol$^{-1}$ at ~20 K before decreasing rapidly at lower temperatures.[2a,4b] The maximum indicates a large ground-state spin ($S$) value, and the low temperature decrease is primarily due to zero-field splitting effects.

To characterize the ground states of **5** and **6**, magnetization (M) data were collected in the 1.0–7.0 T field range and 1.80–4.00 K temperature range to complement the available 10 kG data. Shown in Figures 3 and 4 are plots of the data for **5** and **6**, respectively, as reduced magnetization ($M/N\mu_B$) versus $H/T$, where $N$ is Avogadro's number and $\mu_B$ is the Bohr magneton. For a system occupying only the ground state and experiencing no zero-field splitting (ZFS), the various isofield lines would be superimposed and $M/N\mu_B$ would saturate at a value of $gS$.

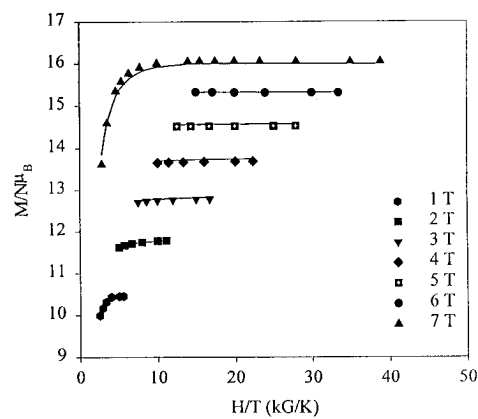

**Figure 3.** Plot of $M/N\mu_B$ vs $H/T$ for complex **5** at the indicated applied fields. The solid lines are fits of the data; see the text for the fit parameters.

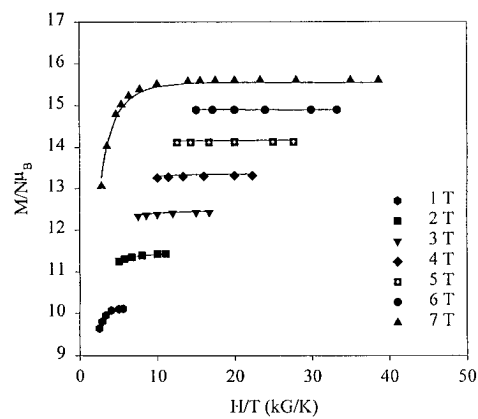

**Figure 4.** Plot of $M/N\mu_B$ vs $H/T$ for complex **6** at the indicated applied fields. The solid lines are fits of the data; see the text for the fit parameters.

The nonsuperimposition of the isofield lines clearly indicates the presence of ZFS. The data were fit assuming only the ground state is populated at these temperatures, using the methods described elsewhere[19,20] involving diagonalization of the spin

---

(19) Vincent, J. B.; Christmas, C.; Chang, H.-R.; Li, Q.; Boyd, P. D. W.; Huffman, J. C.; Hendrickson, D. N.; Christou, G. *J. Am. Chem. Soc.* **1989**, *111*, 2086.

**Table 3.** Selected Bond Distances (Å) and Angles (deg) for Complex **6**·CH$_2$Cl$_2$

| | | | | | | | | | | | |
|---|---|---|---|---|---|---|---|---|---|---|---|
| Mn(1) | O(17) | 1.913(4) | Mn(4) | O(21) | 1.896(4) | Mn(7) | O(15) | 1.914(4) | Mn(10) | O(19) | 1.886(4) |
| Mn(1) | O(18) | 1.891(4) | Mn(4) | O(22) | 1.884(4) | Mn(7) | O(16) | 1.926(4) | Mn(10) | O(25) | 2.198(5) |
| Mn(1) | O(30) | 1.940(4) | Mn(4) | O(43) | 1.927(4) | Mn(7) | O(19) | 1.880(4) | Mn(10) | O(26) | 2.181(4) |
| Mn(1) | O(63) | 1.928(4) | Mn(4) | O(50) | 1.931(4) | Mn(7) | O(20) | 1.850(4) | Mn(10) | O(58) | 1.950(4) |
| Mn(1) | O(76) | 2.192(4) | Mn(4) | O(68) | 2.217(4) | Mn(7) | O(86) | 1.928(4) | Mn(10) | O(65) | 1.942(4) |
| Mn(1) | O(106) | 2.221(5) | Mn(4) | O(94) | 2.244(5) | Mn(8) | O(19) | 1.910(4) | Mn(11) | O(20) | 1.867(4) |
| Mn(2) | O(13) | 1.914(4) | Mn(5) | O(23) | 1.908(4) | Mn(8) | O(20) | 1.890(4) | Mn(11) | O(21) | 1.902(4) |
| Mn(2) | O(14) | 1.904(4) | Mn(5) | O(35) | 1.936(4) | Mn(8) | O(53) | 1.939(4) | Mn(11) | O(48) | 1.933(4) |
| Mn(2) | O(15) | 1.892(4) | Mn(5) | O(40) | 1.938(4) | Mn(8) | O(60) | 1.935(4) | Mn(11) | O(55) | 1.952(4) |
| Mn(2) | O(17) | 1.862(4) | Mn(5) | O(82) | 2.206(5) | Mn(8) | O(88) | 2.203(4) | Mn(11) | O(70) | 2.251(6) |
| Mn(2) | O(18) | 1.860(4) | Mn(5) | O(98) | 2.099(5) | Mn(8) | O(110) | 2.192(4) | Mn(11) | O(112) | 2.164(4) |
| Mn(2) | O(104) | 1.926(4) | Mn(6) | O(14) | 1.903(4) | Mn(9) | O(17) | 1.862(4) | Mn(12) | O(22) | 1.874(4) |
| Mn(3) | O(13) | 1.912(4) | Mn(6) | O(15) | 1.934(4) | Mn(9) | O(24) | 1.879(4) | Mn(12) | O(23) | 1.907(4) |
| Mn(3) | O(14) | 1.925(4) | Mn(6) | O(16) | 1.928(4) | Mn(9) | O(28) | 1.955(5) | Mn(12) | O(27) | 2.224(5) |
| Mn(3) | O(16) | 1.902(4) | Mn(6) | O(23) | 1.880(4) | Mn(9) | O(33) | 1.918(4) | Mn(12) | O(38) | 1.951(4) |
| Mn(3) | O(21) | 1.882(4) | Mn(6) | O(24) | 1.867(4) | Mn(9) | O(74) | 2.097(4) | Mn(12) | O(45) | 1.994(5) |
| Mn(3) | O(22) | 1.853(4) | Mn(6) | O(80) | 1.917(4) | Mn(10) | O(18) | 1.877(4) | Mn(12) | O(100) | 2.131(5) |
| Mn(3) | O(92) | 1.911(4) | Mn(7) | O(13) | 1.897(4) | | | | | | |
| O(17) | Mn(1) | O(18) | 82.32(16) | O(35) | Mn(5) | O(98) | 92.08(21) | O(25) | Mn(10) | O(65) | 85.35(19) |
| O(17) | Mn(1) | O(30) | 95.36(17) | O(40) | Mn(5) | O(82) | 89.38(20) | O(26) | Mn(10) | O(58) | 87.62(19) |
| O(17) | Mn(1) | O(63) | 173.76(18) | O(40) | Mn(5) | O(98) | 93.75(21) | O(26) | Mn(10) | O(65) | 88.49(20) |
| O(17) | Mn(1) | O(76) | 92.38(17) | O(82) | Mn(5) | O(98) | 176.70(18) | O(58) | Mn(10) | O(65) | 82.89(18) |
| O(17) | Mn(1) | O(106) | 85.54(16) | O(14) | Mn(6) | O(15) | 83.06(16) | O(20) | Mn(11) | O(21) | 93.34(16) |
| O(18) | Mn(1) | O(30) | 175.86(18) | O(14) | Mn(6) | O(16) | 84.03(16) | O(20) | Mn(11) | O(48) | 173.96(17) |
| O(18) | Mn(1) | O(63) | 94.51(17) | O(14) | Mn(6) | O(23) | 90.28(16) | O(20) | Mn(11) | O(55) | 90.15(17) |
| O(18) | Mn(1) | O(76) | 92.40(16) | O(14) | Mn(6) | O(24) | 91.46(17) | O(20) | Mn(11) | O(70) | 89.30(17) |
| O(18) | Mn(1) | O(106) | 85.86(16) | O(14) | Mn(6) | O(80) | 174.13(16) | O(20) | Mn(11) | O(112) | 93.52(17) |
| O(30) | Mn(1) | O(63) | 87.47(18) | O(15) | Mn(6) | O(16) | 79.59(15) | O(21) | Mn(11) | O(48) | 92.69(17) |
| O(30) | Mn(1) | O(76) | 91.12(18) | O(15) | Mn(6) | O(23) | 173.28(16) | O(21) | Mn(11) | O(55) | 176.06(17) |
| O(30) | Mn(1) | O(106) | 90.55(18) | O(15) | Mn(6) | O(24) | 95.51(16) | O(21) | Mn(11) | O(70) | 96.15(17) |
| O(63) | Mn(1) | O(76) | 93.12(18) | O(15) | Mn(6) | O(80) | 94.03(16) | O(21) | Mn(11) | O(112) | 93.25(17) |
| O(63) | Mn(1) | O(106) | 88.88(18) | O(16) | Mn(6) | O(23) | 98.90(16) | O(48) | Mn(11) | O(55) | 83.81(18) |
| O(76) | Mn(1) | O(106) | 177.45(17) | O(16) | Mn(6) | O(24) | 173.68(17) | O(48) | Mn(11) | O(70) | 89.89(19) |
| O(13) | Mn(2) | O(14) | 80.46(15) | O(16) | Mn(6) | O(80) | 90.45(17) | O(48) | Mn(11) | O(112) | 86.30(19) |
| O(13) | Mn(2) | O(15) | 83.93(15) | O(23) | Mn(6) | O(24) | 85.51(16) | O(55) | Mn(11) | O(70) | 82.07(18) |
| O(13) | Mn(2) | O(17) | 174.33(17) | O(23) | Mn(6) | O(80) | 92.53(17) | O(55) | Mn(11) | O(112) | 88.34(18) |
| O(13) | Mn(2) | O(18) | 96.47(16) | O(24) | Mn(6) | O(80) | 93.89(18) | O(70) | Mn(11) | O(112) | 170.01(17) |
| O(13) | Mn(2) | O(104) | 92.56(16) | O(13) | Mn(7) | O(15) | 83.81(15) | O(22) | Mn(12) | O(23) | 92.47(16) |
| O(14) | Mn(2) | O(15) | 84.21(15) | O(13) | Mn(7) | O(16) | 83.89(16) | O(22) | Mn(12) | O(27) | 95.64(17) |
| O(14) | Mn(2) | O(17) | 98.02(16) | O(13) | Mn(7) | O(19) | 88.73(16) | O(22) | Mn(12) | O(38) | 173.16(19) |
| O(14) | Mn(2) | O(18) | 174.11(17) | O(13) | Mn(7) | O(20) | 91.63(16) | O(22) | Mn(12) | O(45) | 89.17(18) |
| O(14) | Mn(2) | O(104) | 92.66(17) | O(13) | Mn(7) | O(86) | 176.37(15) | O(22) | Mn(12) | O(100) | 92.36(18) |
| O(15) | Mn(2) | O(17) | 90.49(16) | O(15) | Mn(7) | O(16) | 80.14(15) | O(23) | Mn(12) | O(27) | 95.72(17) |
| O(15) | Mn(2) | O(18) | 90.48(16) | O(15) | Mn(7) | O(19) | 100.29(16) | O(23) | Mn(12) | O(38) | 93.65(18) |
| O(15) | Mn(2) | O(104) | 175.64(17) | O(15) | Mn(7) | O(20) | 173.13(16) | O(23) | Mn(12) | O(45) | 177.90(18) |
| O(17) | Mn(2) | O(18) | 84.55(16) | O(15) | Mn(7) | O(86) | 93.18(16) | O(23) | Mn(12) | O(100) | 92.15(19) |
| O(17) | Mn(2) | O(104) | 92.96(17) | O(16) | Mn(7) | O(19) | 172.52(17) | O(27) | Mn(12) | O(38) | 80.73(20) |
| O(18) | Mn(2) | O(104) | 92.50(17) | O(16) | Mn(7) | O(20) | 94.30(16) | O(27) | Mn(12) | O(45) | 85.42(20) |
| O(13) | Mn(3) | O(14) | 79.98(15) | O(16) | Mn(7) | O(86) | 93.61(16) | O(27) | Mn(12) | O(100) | 168.51(19) |
| O(13) | Mn(3) | O(16) | 84.15(16) | O(19) | Mn(7) | O(20) | 84.70(16) | O(38) | Mn(12) | O(45) | 84.79(19) |
| O(13) | Mn(3) | O(21) | 98.96(16) | O(19) | Mn(7) | O(86) | 93.82(16) | O(38) | Mn(12) | O(100) | 90.44(21) |
| O(13) | Mn(3) | O(22) | 173.54(16) | O(20) | Mn(7) | O(86) | 91.18(16) | O(45) | Mn(12) | O(100) | 86.46(21) |
| O(13) | Mn(3) | O(92) | 92.61(16) | O(19) | Mn(8) | O(20) | 82.81(16) | Mn(2) | O(13) | Mn(3) | 99.82(16) |
| O(14) | Mn(3) | O(16) | 84.13(16) | O(19) | Mn(8) | O(53) | 172.58(17) | Mn(2) | O(13) | Mn(7) | 95.16(16) |
| O(14) | Mn(3) | O(21) | 174.53(16) | O(19) | Mn(8) | O(60) | 95.29(16) | Mn(3) | O(13) | Mn(7) | 95.47(17) |
| O(14) | Mn(3) | O(22) | 95.41(16) | O(19) | Mn(8) | O(88) | 86.31(16) | Mn(2) | O(14) | Mn(3) | 99.71(16) |
| O(14) | Mn(3) | O(92) | 91.68(16) | O(19) | Mn(8) | O(110) | 95.32(16) | Mn(2) | O(14) | Mn(6) | 95.82(17) |
| O(16) | Mn(3) | O(21) | 90.43(16) | O(20) | Mn(8) | O(53) | 95.33(16) | Mn(3) | O(14) | Mn(6) | 94.99(16) |
| O(16) | Mn(3) | O(22) | 90.91(16) | O(20) | Mn(8) | O(60) | 177.57(17) | Mn(2) | O(15) | Mn(6) | 95.20(16) |
| O(16) | Mn(3) | O(92) | 175.09(17) | O(20) | Mn(8) | O(88) | 85.66(15) | Mn(2) | O(15) | Mn(7) | 95.34(16) |
| O(21) | Mn(3) | O(22) | 85.19(16) | O(20) | Mn(8) | O(110) | 91.09(16) | Mn(6) | O(15) | Mn(7) | 100.23(17) |
| O(21) | Mn(3) | O(92) | 93.73(16) | O(53) | Mn(8) | O(60) | 86.37(16) | Mn(3) | O(16) | Mn(6) | 94.97(16) |
| O(22) | Mn(3) | O(92) | 92.04(17) | O(53) | Mn(8) | O(88) | 86.39(16) | Mn(3) | O(16) | Mn(7) | 94.81(17) |
| O(21) | Mn(4) | O(22) | 83.95(16) | O(53) | Mn(8) | O(110) | 91.89(17) | Mn(6) | O(16) | Mn(7) | 100.02(17) |
| O(21) | Mn(4) | O(43) | 173.76(19) | O(60) | Mn(8) | O(88) | 92.71(16) | Mn(1) | O(17) | Mn(2) | 95.40(16) |
| O(21) | Mn(4) | O(50) | 96.76(17) | O(60) | Mn(8) | O(110) | 90.59(16) | Mn(1) | O(17) | Mn(9) | 124.35(20) |
| O(21) | Mn(4) | O(68) | 92.38(15) | O(88) | Mn(8) | O(110) | 176.17(16) | Mn(2) | O(17) | Mn(9) | 133.43(22) |
| O(21) | Mn(4) | O(94) | 86.61(16) | O(17) | Mn(9) | O(24) | 94.63(16) | Mn(1) | O(18) | Mn(2) | 96.20(16) |
| O(22) | Mn(4) | O(43) | 92.26(17) | O(17) | Mn(9) | O(28) | 90.48(18) | Mn(1) | O(18) | Mn(10) | 129.86(21) |

**Table 3** (Continued)

| | | | | | | | | | | | |
|---|---|---|---|---|---|---|---|---|---|---|---|
| O(22) | Mn(4) | O(50) | 178.64(18) | O(17) | Mn(9) | O(33) | 171.92(19) | Mn(2) | O(18) | Mn(10) | 132.28(21) |
| O(22) | Mn(4) | O(68) | 88.51(17) | O(17) | Mn(9) | O(74) | 94.51(16) | Mn(7) | O(19) | Mn(8) | 94.61(17) |
| O(22) | Mn(4) | O(94) | 84.84(16) | O(24) | Mn(9) | O(28) | 169.23(19) | Mn(7) | O(19) | Mn(10) | 131.68(20) |
| O(43) | Mn(4) | O(50) | 86.93(19) | O(24) | Mn(9) | O(33) | 91.42(17) | Mn(8) | O(19) | Mn(10) | 126.55(19) |
| O(43) | Mn(4) | O(68) | 92.47(19) | O(24) | Mn(9) | O(74) | 95.25(18) | Mn(7) | O(20) | Mn(8) | 96.31(17) |
| O(43) | Mn(4) | O(94) | 88.12(19) | O(28) | Mn(9) | O(33) | 82.70(18) | Mn(7) | O(20) | Mn(11) | 134.65(20) |
| O(50) | Mn(4) | O(68) | 92.62(18) | O(28) | Mn(9) | O(74) | 93.79(20) | Mn(8) | O(20) | Mn(11) | 126.32(19) |
| O(50) | Mn(4) | O(94) | 94.03(18) | O(33) | Mn(9) | O(74) | 90.27(19) | Mn(3) | O(21) | Mn(4) | 94.07(16) |
| O(68) | Mn(4) | O(94) | 173.34(16) | O(18) | Mn(10) | O(19) | 94.49(16) | Mn(3) | O(21) | Mn(11) | 131.55(21) |
| O(23) | Mn(5) | O(24) | 83.59(16) | O(18) | Mn(10) | O(25) | 89.52(18) | Mn(4) | O(21) | Mn(11) | 122.92(20) |
| O(23) | Mn(5) | O(35) | 173.68(21) | O(18) | Mn(10) | O(26) | 92.86(16) | Mn(3) | O(22) | Mn(4) | 95.41(16) |
| O(23) | Mn(5) | O(40) | 94.62(17) | O(18) | Mn(10) | O(58) | 172.86(18) | Mn(3) | O(22) | Mn(12) | 133.31(21) |
| O(23) | Mn(5) | O(82) | 86.66(16) | O(18) | Mn(10) | O(65) | 90.00(17) | Mn(4) | O(22) | Mn(12) | 130.20(21) |
| O(23) | Mn(5) | O(98) | 94.13(18) | O(19) | Mn(10) | O(25) | 94.43(17) | Mn(5) | O(23) | Mn(6) | 94.33(17) |
| O(24) | Mn(5) | O(35) | 95.23(18) | O(19) | Mn(10) | O(26) | 91.53(18) | Mn(5) | O(23) | Mn(12) | 122.10(19) |
| O(24) | Mn(5) | O(40) | 175.57(20) | O(19) | Mn(10) | O(58) | 92.63(17) | Mn(6) | O(23) | Mn(12) | 131.96(20) |
| O(24) | Mn(5) | O(82) | 86.47(17) | O(19) | Mn(10) | O(65) | 175.51(16) | Mn(5) | O(24) | Mn(6) | 94.74(18) |
| O(24) | Mn(5) | O(98) | 90.42(18) | O(25) | Mn(10) | O(26) | 173.40(19) | Mn(5) | O(24) | Mn(9) | 126.85(20) |
| O(35) | Mn(5) | O(40) | 86.11(19) | O(25) | Mn(10) | O(58) | 89.27(21) | Mn(6) | O(24) | Mn(9) | 133.37(21) |
| O(35) | Mn(5) | O(82) | 87.07(20) | | | | | | | | |

**Table 4.** Comparison of $Mn^{III}$–O Bond Lengths (Å) in $[Mn_{12}O_{12}(O_2CR)_8(O_2CR')_8(H_2O)_x]$ ($x$ = 3 or 4) Complexes

| complex | $Mn^{III}$–O(eq)$^a$ | $Mn^{III}$–O(ax)$^a$ | Δ |
|---|---|---|---|
| R = R' = Me (**1**) | 1.928(67) | 2.176(44) | 0.248 |
| R = R' = CH$_2$Bu$^t$ (**2**) | 1.923(62) | 2.185(85) | 0.262 |
| R = R' = Et (**3**) | 1.920(61) | 2.166(59) | 0.246 |
| R = CHCl$_2$, R' = CH$_2$Bu$^t$ (**5**) | 1.918(48) | 2.193(90) | 0.275 |
| R = CHCl$_2$, R' = Et (**6**) | 1.916(81) | 2.202(105) | 0.286 |

$^a$ Average values. Numbers in parentheses are the greatest deviation from the mean; eq = equatorial, ax = axial.

Hamiltonian matrix including axial ZFS ($D\hat{S}_z^2$) and Zeeman interactions and incorporating a full powder average of the magnetization. The best fits are shown as solid lines in Figures 3 and 4, and the fitting parameters were $S = 10$, $g = 1.89$, and $D = -0.45$ cm$^{-1}$ = $-0.65$ K for complex **5**, and $S = 10$, $g = 1.83$, and $D = -0.42$ cm$^{-1}$ = $-0.60$ K for complex **6**. These values are typical of the Mn$_{12}$ family and demonstrate that the presence of two types of carboxylate ligands with distinctly different basicities does not significantly perturb the properties of the Mn$_{12}$ complexes. To probe whether it might, however, influence the rate of magnetization relaxation, ac susceptibility studies were performed.

**Alternating Current Susceptibility Studies.** In an ac susceptibility experiment, a weak field (typically 1–5 Oe) oscillating at a particular frequency is applied to a sample to probe the dynamics of its magnetization relaxation.[2,4,21] An out-of-phase ac susceptibility signal ($\chi_M''$) is observed when the rate at which the magnetization (magnetic moment) of a molecule (or collection of molecules) relaxes (reorients) is close to the operating frequency of the ac field. Thus, if a collection of SMMs is maintained at a certain temperature and the frequency of the ac magnetic field is varied, a maximum in the $\chi_M''$ signal will occur when the oscillation frequency of the field equals the rate at which a molecule can interconvert between the halves of the potential energy double well shown in Figure 5. This figure shows a plot of the potential energy of an $S = 10$ molecule exhibiting easy-axis type anisotropy (negative $D$ value)

(20) Hendrickson, D. N.; Christou, G.; Schmitt, E. A.; Libby, E.; Bashkin, J. S.; Wang, S.; Tsai, H.-L.; Vincent, J. B.; Boyd, P. D. W.; Huffman, J. C.; Folting, K.; Li, Q.; Streib, W. E. *J. Am. Chem. Soc.* **1992**, *114*, 2455.

(21) Novak, M. A.; Sessoli, R. In *Quantum Tunneling of Magnetization— QTM '94*; Gunther, L., Barbara, B., Eds.; Kluwer: Amsterdam, 1995; pp 171–188.

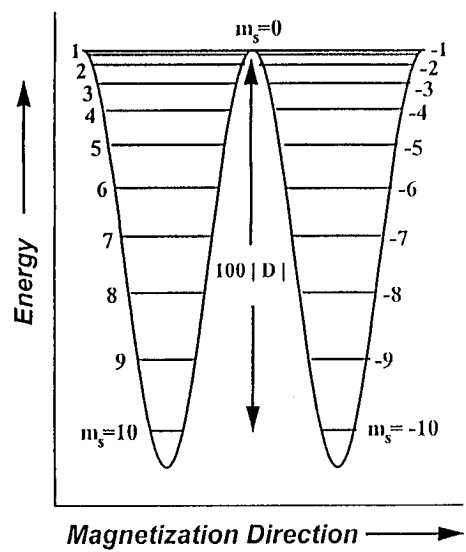

**Figure 5.** Double-well potential energy plot for an $S = 10$ molecule with axial symmetry and easy-axis type anisotropy. The energy separation between the $M_s = \pm 10$ and $M_s = 0$ orientations of the magnetization is given by $S^2|D|$.

as its magnetization (magnetic moment) vector changes from spin "up" ($M_s = -10$) to spin "down" ($M_s = +10$) via intermediate orientations. The potential energy barrier ($U$) is given by $U = S^2|D|$ for integer-spin systems with axial symmetry. Frequency-dependent $\chi_M''$ signals have been observed in the ac susceptibility studies of all Mn$_{12}$ SMMs and are considered a diagnostic signature of the SMM property.[2,4,20,21] In addition, ac magnetic susceptibility data can be employed to obtain the effective energy barrier ($U_{eff}$) for magnetization relaxation and even the spin of the ground state.[21] For these reasons, ac susceptibility studies were performed on complexes **5** and **6**.

The ac susceptibilities of **5** and **6** (Figures 6 and 7, respectively) were recorded in a 3.5 G ac field oscillating at various frequencies in the range indicated. The upper panels show the $\chi_M'T$ versus $T$ plots, where $\chi_M'$ is the in-phase magnetic susceptibility, and the lower panels show the $\chi_M''$ versus $T$ plots. The frequency-dependent decrease in the $\chi_M'T$ versus $T$ plots is a signature of the magnetization relaxation rate becoming comparable with the ac frequency. Consistent with this is the appearance of $\chi_M''$ signals in the 4–7 K range, which confirms that **5** and **6** are displaying the superparamagnetic-like properties of SMMs. In both Figures 6 and 7, it is clear that only one $\chi_M''$ peak is seen at each frequency, and these samples therefore do not exhibit the new phenomenon of Jahn−Teller isomerism,[22] whereby some $Mn_{12}$ molecules possess an abnormal orientation of one JT axis. The latter leads to faster relaxation rates and $\chi_M''$ peaks at correspondingly lower temperatures (2−4 K). Also note that the $\chi_M'T$ value at its near plateau value at ∼10 K of 50−52 cm$^3$ K mol$^{-1}$ for **5** and **6** can be used to determine the ground-state $S$ values of the complexes, assuming only the ground state is occupied at this temperature. A $\chi_M'T$ value of ∼51 cm$^3$ K mol$^{-1}$ corresponds to an $S = 10$ system with $g = 1.93$, consistent with the dc magnetization results above.

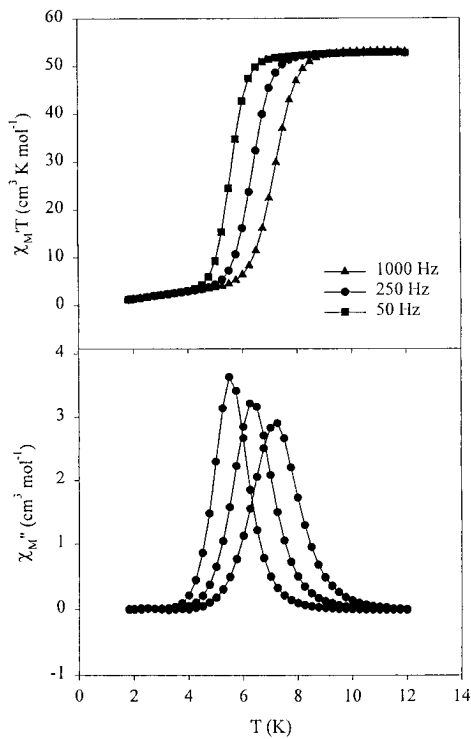

**Figure 6.** In-phase ($\chi_M'$) and out-of-phase ($\chi_M''$) ac susceptibility signals for [$Mn_{12}O_{12}(O_2CCHCl_2)_8(O_2CCH_2Bu^t)_8(H_2O)_3$] (**5**) at the indicated frequencies.

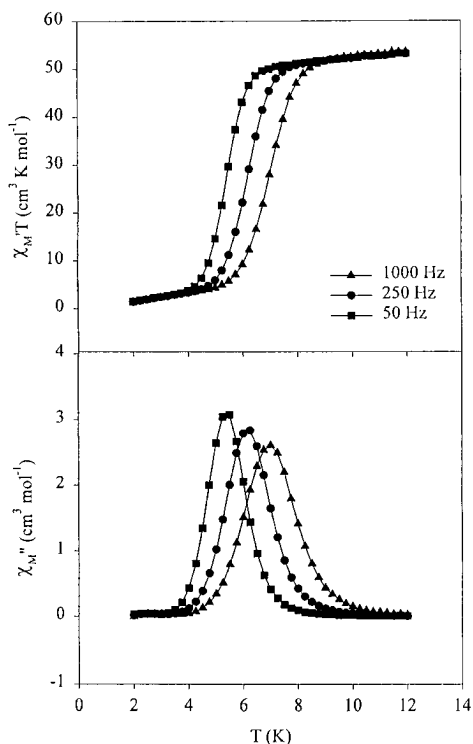

**Figure 7.** In-phase ($\chi_M'$) and out-of-phase ($\chi_M''$) ac susceptibility signals for [$Mn_{12}O_{12}(O_2CCHCl_2)_8(O_2CEt)_8(H_2O)_3$] (**6**) at the indicated frequencies.

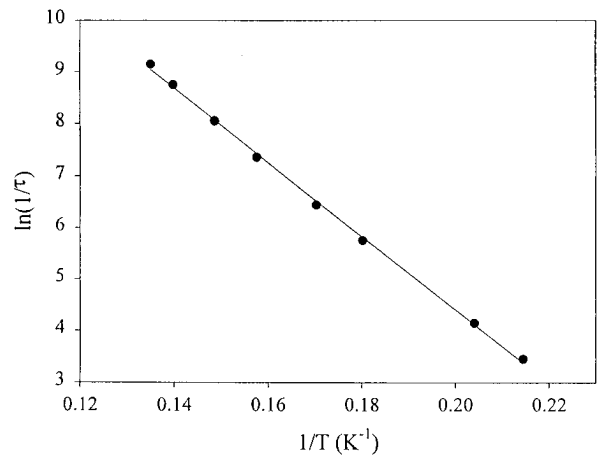

**Figure 8.** Plot of the natural logarithm of relaxation rate, $\ln(1/\tau)$, vs inverse temperature for [$Mn_{12}O_{12}(O_2CCHCl_2)_8(O_2CCH_2Bu^t)_8(H_2O)_3$] (**5**) using $\chi_M''$ vs $T$ data at different frequencies. The solid line is a fit to the Arrhenius equation; see the text for the fit parameters.

At the temperature of the $\chi_M''$ versus $T$ peak maximum, the relaxation rate ($1/\tau$) equals the ac frequency ($\omega$).[21] Thus, the $\chi_M''$ versus $T$ plots at different frequencies provide $1/\tau$ versus $T$ data, and a kinetic analysis can be performed using the Arrhenius relationship (eq 5). This is the

$$\ln(1/\tau) = -U_{eff}/kT + \ln(1/\tau_0) \quad (5)$$

characteristic behavior of a thermally activated Orbach process,[23] where $U_{eff}$ is the effective anisotropy energy barrier, $k$ is the Boltzmann constant, and $1/\tau_0$ is the pre-exponential term. Plots of $\ln(1/\tau)$ versus $1/T$ for **5** and **6** using $\chi_M''$ versus $T$ data are shown in Figures 8 and 9, respectively, with the least-squares fit to eq 4 shown as a solid line. From the slope and intercept, it was determined that $U_{eff} = 50$ cm$^{-1} = 72$ K and $1/\tau_0 = 1.3 \times 10^8$ s$^{-1}$ for complex **5**, and $U_{eff} = 49$ cm$^{-1} = 71$ K and $1/\tau_0 = 1.6 \times 10^8$ s$^{-1}$ for complex **6**. These $U_{eff}$ values both fall in the range 42−50 cm$^{-1}$ (60−72 K) observed previously for [$Mn_{12}O_{12}(O_2CR)_{16}(H_2O)_4$] complexes.[2b,21,22]

The $U_{eff}$ values for **5** and **6** obtained from Arrhenius plots may be compared with $U$ values calculated as $S^2|D|$, which are 45 cm$^{-1}$ (65 K) and 42 cm$^{-1}$ (60 K) for **5** and **6**, respectively, using $S = 10$ and the $D$ values obtained from the $M/N\mu_B$ versus $H/T$ fits in Figures 3 and 4. In both cases, $U_{eff} > U$ by about 10−20%. In [$Mn_{12}O_{12}(O_2CMe)_{16}(H_2O)_4$]·2MeCO$_2$H·4H$_2$O (**1**·2MeCO$_2$H·4H$_2$O), $U_{eff}$ is 64 K, and $S^2|D| = 72$ K using $S = 10$ and $D = -0.5$ cm$^{-1} = -0.72$ K. Thus, $U_{eff} < U$, which is rationalized to be due to the presence of quantum tunneling of the magnetization[21,24,25] (QTM), leading to a smaller effective

(22) (a) Sun, Z.; Ruiz, D.; Dilley, N. R.; Soler, M.; Ribas, J.; Folting, K.; Maple, M. B.; Christou, G.; Hendrickson, D. *Chem. Commun.* **1999**, 1973. (b) Aubin, S. M. J.; Sun, Z.; Eppley, H. J.; Rumberger, E.; Guzei, I. A.; Folting, K.; Gantzel, P. K.; Rheingold, A. L.; Christou, G.; Hendrickson, D. N. *Inorg. Chem.*, manuscript in press.
(23) (a) van Duyneveldt, A. J. In *Magnetic Molecular Materials*; Gatteschi, D., Kahn, O., Miller, J., Palacio, F., Eds.; Kluwer Academic Publishers: London, 1991. (b) Abram, A.; Bleaney, B. *Electron Paramagnetic Resonance of Transition Ions*; Dover Press: Mineola, NY, 1986.
(24) Friedman, J. R.; Sarachik, M. P.; Tejada, J.; Maciejewski, J.; Ziolo, R. *J. Appl. Phys.* **1996**, *79*, 6031.

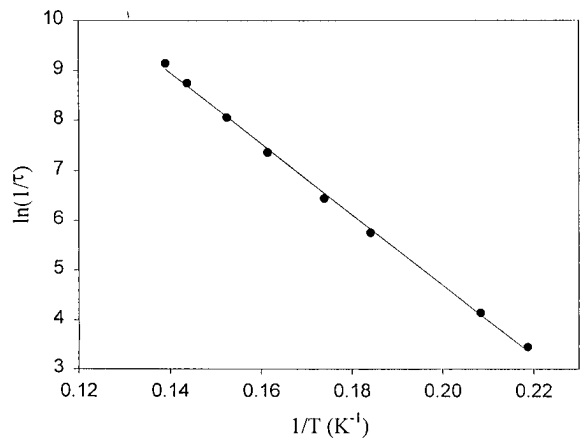

**Figure 9.** Plot of the natural logarithm of relaxation rate, $\ln(1/\tau)$, vs inverse temperature for $[Mn_{12}O_{12}(O_2CCHCl_2)_8(O_2CEt)_7(H_2O)_3]$ (**6**) using $\chi_M''$ vs $T$ data at different frequencies. The solid line is a fit to the Arrhenius equation; see the text for the fit parameters.

barrier to relaxation ($U_{eff}$) than that calculated for a purely thermally activated process ($U$). In contrast, we have found $U_{eff} > U$ values for **5** and **6**. We also obtained $U_{eff} > U$ for $[Mn_{12}O_{12}(NO_3)_4(O_2CCH_2Bu^t)_{16}(H_2O)_4]$[14] and, on other occasions, have also observed $U_{eff} > U$ or $U_{eff} \approx U$ in $Mn_{12}$ complexes. We believe the origin of this effect lies in the different site symmetries of the $Mn_{12}$ complexes in **1**·2MeCO$_2$H·4H$_2$O ($S_4$, i.e., axial symmetry) versus **5**, **6**, and other $Mn_{12}$ complexes ($C_1$, i.e., rhombic symmetry). We suspect the origin of the discrepancy may be in the transverse zero-field interaction terms present in the spin Hamiltonian of lower symmetry $Mn_{12}$ complexes but absent in the axial symmetry of complex **1**. The fitting of the $M/N\mu_B$ versus $H/T$ data for **5** and **6** includes only the axial zero-field parameter $D$, and axial symmetry is assumed in using the relationship $U = S^2|D|$. Inclusion of additional fitting parameters into the $M/N\mu_B$ versus $H/T$ fits is unjustified, given the number of parameters already employed ($S$, $g$, $D$). We conclude that the $D$ values obtained for **5** and **6** are subject to some uncertainty and that our data do not allow for an accurate evaluation of $U$. In contrast, $U_{eff}$ is considered much more reliable. Further study is in progress on this matter.

**$^1$H NMR Spectroscopy.** The crystal structures of **5** and **6** clearly demonstrate that in the solid state the two types of carboxylate groups are localized at either equatorial or axial positions. To probe whether this situation persists in solution, $^1$H NMR spectra of **5** and **6** were investigated in CD$_2$Cl$_2$ and compared with those of the parent homocarboxylate complexes **2**−**4**. In Figure 10 are shown the spectra of **2**, **4**, and **5**; the peak assignments for **2** and **4** are based on detailed variable temperature (VT) studies and $T_1$ measurements described elsewhere.[14] At room temperature, there is a fluxional process that rapidly exchanges the H$_2$O molecules and the RCO$_2^-$ groups that have both the O atoms on the JT elongation axes. Thus, the Mn$_{12}$ complexes exhibit effective $D_{2d}$ symmetry, giving two types of axial RCO$_2^-$ groups and one type of equatorial group, in a 1:1:2 relative integration ratio, and three resonances are seen in the spectrum of **4** (Figure 10, bottom). The spectrum of **2** (Figure 10, top) shows the same 1:1:2 pattern for the Bu$^t$ groups in the $\delta = 0-6$ ppm range. However, four resonances are seen in a 1:1:1:1 ratio for the CH$_2$ groups, but this is as expected because the equatorial Bu$^t$CH$_2$CO$_2^-$ groups have diastereotopic CH$_2$ hydrogen nuclei in $D_{2d}$ symmetry. The

---

(25) Friedman, J. R.; Sarachik, M. P.; Tejada, J.; Ziolo, R. *Phys. Rev. Lett.* **1996**, *76*, 3830.

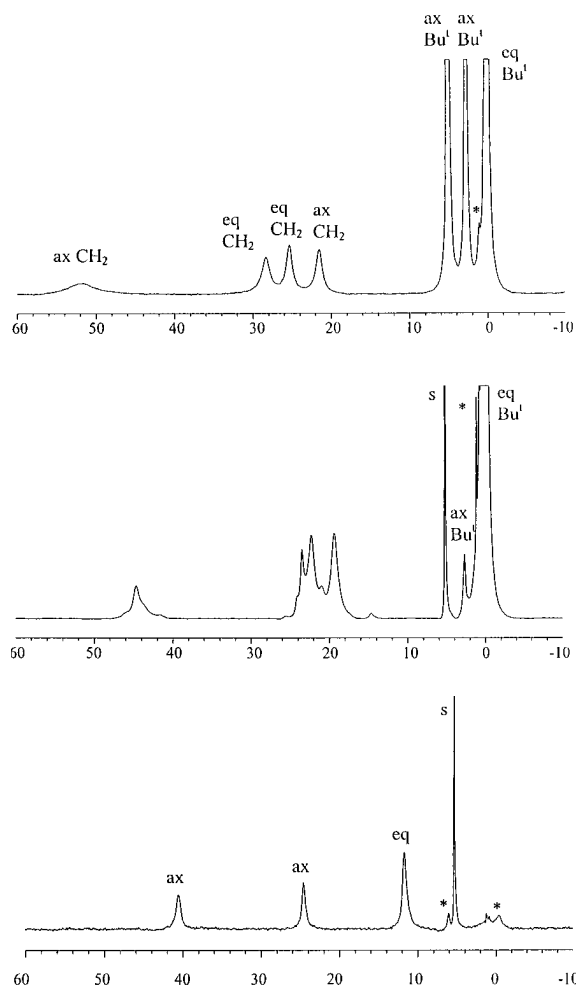

**Figure 10.** $^1$H NMR spectra in CD$_2$Cl$_2$ at ∼23 °C of (top) $[Mn_{12}O_{12}(O_2CCH_2Bu^t)_{16}(H_2O)_4]$ (**2**), (middle) $[Mn_{12}O_{12}(O_2CCHCl_2)_8(O_2CCH_2Bu^t)_8(H_2O)_4]$ (**5**), and (bottom) $[Mn_{12}O_{12}(O_2CCHCl_2)_{16}(H_2O)_4]$ (**4**); ax = axial, eq = equatorial, s = CHDCl$_2$, * = H$_2$O and other solvent impurities.

spectrum of the mixed-carboxylate species **5** (Figure 10, middle) shows more than the five resonances that would be expected for pure $[Mn_{12}O_{12}(O_2CCHCl_2)_8(O_2CCH_2Bu^t)_8(H_2O)_4]$ with effective $D_{2d}$ solution symmetry, two axial CHCl$_2$ resonances, two equatorial (diastereotopic) CH$_2$ resonances, and one equatorial Bu$^t$ resonance. Clearly, some level of ligand redistribution is suggested, because at least 12 resonances are visible. The Bu$^t$ resonances are particularly informative: in addition to the intense equatorial Bu$^t$ resonance, there is a signal at $\delta \approx 3$ ppm from an axial Bu$^t$ group. In addition, the solvent signal at $\delta = 5.32$ ppm probably also is coincidental with the resonance of the second type of axial Bu$^t$ groups. We conclude that carboxylate exchange between axial and equatorial sites is occurring to give species of lower symmetry and a resulting increased number of NMR resonances. Indeed, both intramolecular and intermolecular processes are likely occurring, as summarized in eqs 6 and 7. VT studies, as well as an investigation of the spectra obtained from the reaction of **2** and **4** at different ratios, are required to probe the processes further, but note that the formation of **5** and **6** from the homocarboxylate starting materials likely requires the same kind of intra- and intermolecular processes. Clearly, however, the main species in solution is **5**, given that the axial $Bu^t$ resonances are so relatively weak compared with the equatorial $Bu^t$ resonance, and it is therefore not surprising that this species is the one obtained in the solid state (assuming the solubilities of the various species are comparable). The spectra of **3**, **4**, and **6** were similarly compared, and the same general behavior is observed; the spectrum of **6** is consistent with more than just $[Mn_{12}O_{12}(O_2CHCl_2)_8(O_2CEt)_8(H_2O)_4]$ being present.

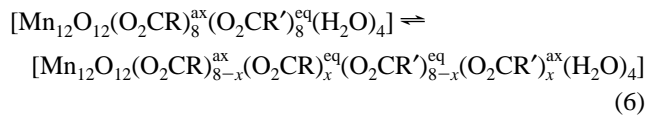

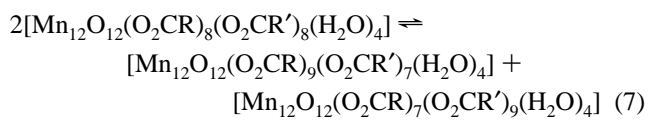

## Conclusions

The reaction between molar equivalents of two different $[Mn_{12}O_{12}(O_2CR)_{16}(H_2O)_4]$ complexes is a convenient method for preparing mixed-carboxylate $[Mn_{12}O_{12}(O_2CR)_8(O_2CR')_8(H_2O)_4]$ complexes. The two types of carboxylate groups are ordered in axial and equatorial sites. This synthetic method is favored over the reaction of $[Mn_{12}O_{12}(O_2CR)_{16}(H_2O)_4]$ with $R'CO_2H$, because it avoids the presence of free $RCO_2H$ and $R'CO_2H$ groups in solution that are more likely to lead to multiple species in equilibrium. Structurally, the complexes are identical to $[Mn_{12}O_{12}(O_2CR)_{16}(H_2O)_4]$ species vis-à-vis the $[Mn_{12}O_{12}]$ core, disposition of JT axes, and so forth, and this is also reflected in the magnetic properties that reveal retention of slow relaxation of magnetization and the resulting single-molecule magnetism properties.

The present work augurs well for efforts to bind these complexes to other molecules or to surfaces. The ordered ligand nature of the mixed-carboxylate complexes represents a way of site-specific introduction of specific types of ligands, such as bifunctional carboxylates $O_2C(CH_2)_nX$ whose X groups are themselves metal-binding units, or $RCO_2^-$ groups with particular types of R groups to enhance shape anisotropy or other properties of the $Mn_{12}$ complexes. Such studies are currently in progress.

**Acknowledgment.** This work was supported by National Science Foundation grants to D.N.H. and G.C.

**Supporting Information Available:** X-ray crystallographic files in CIF format for complexes **5** and **6**. This material is available free of charge via the Internet at http://pubs.acs.org.

IC0104048